\documentclass[aps,prl,draft,showpacs,twocolumn]{revtex4}

\newcommand{\tr}{\mbox{Tr} }
\newcommand{\ket}[1]{\left | #1 \right \rangle}

\begin{document}
\title{The Bound of Entanglement of Superpositions with More Than Two Components}

\author{Yang Xiang}
\email{njuxy@sina.com}
\author{Shi-Jie Xiong}
\author{Fang-Yu Hong }
\affiliation{National Laboratory of Solid State Microstructures and
Department of Physics, \\Nanjing University, Nanjing 210093, China}
\date{\today}

\begin{abstract}
A bipartite quantum state (for two systems in any dimensions) can be
decomposed as a superposition of many components. For a
superposition of more than two components we prove that there is a
bound of the entanglement of the superposition state which can be
expressed according to entanglements of its component states.
Especially, if the component states are mutually bi-orthogonal, the
entanglement of the superposition state can be exactly given in
terms of the entanglements of the states being superposed.
\end{abstract}

\pacs{03.67.-a, 03.65.Ta, 03.65.Ud}

\maketitle


Superposition and entanglement are two bases of quantum mechanics.
For a bipartite pure state, Popescu and Rohrlich \cite{Popescu}
proved that there is a unique measure of the entanglement of it, the
von Neumann entropy of the reduced state of either of the parties
\cite{bennett}. For example, the entanglement of the bipartite pure
state $|\psi\rangle$ is :
\begin{eqnarray}
E(\psi)\equiv
S(\tr_A|\psi\rangle\langle\psi|)&=&S(\tr_B|\psi\rangle\langle\psi|),
\end{eqnarray}
where the von Neumann entropy is defined as
\begin{eqnarray}
S(\rho)=-\tr(\rho\log\rho).
\end{eqnarray}
In this paper $\log$ denotes $\log_{2}$.

As Linden {\it et al.} pointed out \cite{linden}, entanglement is a
global property of a state and it originates from the superposition
of different components. For example, four Bell kets are bipartite
pure superposition states and each of them is maximally entangled,
but every component in the superpositions is unentangled as it can
be expressed by a direct product of pure quantum states of the
parties. Inversely we can pick out two proper Bell kets to compose a
superposition state which is unentangled. In other cases, the forms
of two states are almost the same but they do not necessarily have
nearly the same entanglement. Thus, there may be some implicit
relations between the superposition and the entanglement. We can
raise a problem: Given a bipartite superposition state, what is the
relation between the entanglement of it and those of the components
in the superposition? Linden {\it et al.} \cite{linden} discussed a
state $|\Gamma\rangle$ of two parties and presented a certain
decomposition of it as a superposition of two terms
$|\Gamma\rangle=\alpha|\psi\rangle+\beta|\phi\rangle$ with
$|\alpha|^2+|\beta|^2=1$. They found an upper bound of the
entanglement of $|\Gamma\rangle$ in terms of the entanglements of
$|\psi\rangle$ and $|\phi\rangle$. Subsequently, several authors
\cite{yu,ou,niset,Cavalcanti,song,gour} have discussed some related
problems about this issue: in \cite{yu,ou,niset}, the authors
generalized this result to include different measures of
entanglement; in \cite{Cavalcanti,song}, the authors discussed the
entanglement of superpositions of multipartite states; in
\cite{gour}, the author found tight lower and upper bounds on the
entanglement of a superposition of two bipartite states.

In this paper, we generalize the conclusion to a pure bipartite
state $|\Psi\rangle$ which is a superposition of more than two
$(n>2)$ components
\begin{eqnarray}
|\Psi\rangle&=&\alpha_1|\phi_1\rangle+\alpha_2
|\phi_2\rangle+\cdot\cdot\cdot+\alpha_n|\phi_n\rangle , \label{3}
\end{eqnarray}
where $\phi_1,\phi_2\cdot\cdot\cdot\phi_n$ are nonorthogonal and
normalized, $\alpha_1,\alpha_2\cdot\cdot\cdot\alpha_n$ satisfy
$\sum\limits_{i = 1}^n {N_i ^2 |\alpha _i |^2 }  = 1$ with $N_i$s
being coefficients depending on $n$ as will be discussed below. We
present an upper bound on the entanglement of $\Psi$ given by the
inequality
\begin{widetext}
\begin{eqnarray}
\|\alpha_1|\phi_1\rangle+\alpha_2|\phi_2\rangle+\cdot\cdot\cdot+\alpha_n|\phi_n\rangle\|^2
\cdot
E(\alpha_1\phi_1+\alpha_2\phi_2+\cdot\cdot\cdot+\alpha_n\phi_n)\leq
\sum\limits_{i = 1}^n {N_i ^2 |\alpha _i |^2 E(\phi_i)}+h_n ,
\label{xymain-ineq}
\end{eqnarray}
\end{widetext}
where the notation
$E(\alpha_1\phi_1+\alpha_2\phi_2+\cdot\cdot\cdot+\alpha_n\phi_n)$
denotes the entanglement of the normalized version of the state
$\alpha_1|\phi_1\rangle+\alpha_2|\phi_2\rangle+\cdot\cdot\cdot+\alpha_n|\phi_n\rangle$,
and $h_n=-\sum\limits_{i = 1}^n {N_i ^2 |\alpha _i |^2 \log(N_i ^2
|\alpha _i |^2)}$.

Before embarking on our study, it is worth introducing an inequality
first which will be used repeatedly in the following. For a mixed
state $\sum\limits_{i = 1}^n {p_i\rho_i}$, the von Neumann entropy
satisfies \cite{nielsen}
\begin{eqnarray}
\sum\limits_{i=1}^n {p_i S(\rho_i)}\leq S(\sum\limits_{i=1}^n
{p_i\rho_i})\leq\sum\limits_{i=1}^n {p_i S(\rho_i)}+H
\label{assis-ineq},
\end{eqnarray}
where $H=-\sum\limits_{i=1}^n {p_i\log\rho_i}$.



In the case that the component states
$\phi_1,\phi_2,\cdot\cdot\cdot,\phi_n$ are mutually biorthogonal we
can easily present an exact expression of the entanglement of $\Psi$
in terms of the entanglements of
$\phi_1,\phi_2,\cdot\cdot\cdot,\phi_n$\cite{xiang}:
\begin{eqnarray}
E(\Psi)&=&E(\alpha_{1}\phi_{1}+\alpha_{2}\phi_{2}+\cdot\cdot\cdot+\alpha_{n}\phi_{n})\nonumber\\
&=&|\alpha_{1}|^2E(\phi_{1})+|\alpha_{2}|^2E(\phi_{2})+\cdot\cdot\cdot\nonumber\\
&
&+|\alpha_{n}|^2E(\phi_{n})+h_{n}^{'}(\alpha_{1},\alpha_{2},\cdot\cdot\cdot,\alpha_{n}),
\label{ans1}
\end{eqnarray}
where
\begin{eqnarray}
h_{n}^{'}(\alpha_{1},\alpha_{2},\cdot\cdot\cdot,\alpha_{n})
&=&-\sum\limits_{i=1}^{n}{|\alpha_{i}|^2\log{|\alpha_{i}|^2}}.
\end{eqnarray}
The definiens of biorthogonal states is that if any two states
$\phi_i,\phi_j$ ($i\neq j$) in
$\phi_1,\phi_2,\cdot\cdot\cdot,\phi_n$ satisfy
\begin{eqnarray}
\tr_A[\tr_B(|\phi_i\rangle\langle\phi_i|)\tr_B(|\phi_j\rangle\langle\phi_j|)]
&=&0, \nonumber\\
\tr_B[\tr_A(|\phi_i\rangle\langle\phi_i|)\tr_A(|\phi_j\rangle\langle\phi_j|)&=&0,
\end{eqnarray}
we say that $\phi_1,\phi_2\cdot\cdot\cdot\phi_n$ are mutually
biorthogonal, or say that they are biorthogonal states.

%
%
%
%
%
%
%
%
%
%
%
Now we discuss the general case that
$\phi_1,\phi_2,\cdot\cdot\cdot,\phi_n$ are nonorthogonal and provide
a proof of inequality (\ref{xymain-ineq}). The case that
$\phi_1,\phi_2,\cdot\cdot\cdot,\phi_n$ are orthogonal (but not
biorthogonal) is just a special situation of the general case. At
the end we will give another version of (\ref{xymain-ineq}) for an
arbitrary superposition state
$|\Psi\rangle=\alpha_1|\phi_1\rangle+\alpha_2|\phi_2\rangle+
\cdot\cdot\cdot+\alpha_n|\phi_n\rangle$, i.e., without the
constrained condition $\sum\limits_{i = 1}^n {N_i ^2 |\alpha _i |^2
} =1$.

As considered in \cite{linden}, we think that Alice has a
$n$-dimensional Hilbert space $\mathcal {H}_{a}$ besides Hilbert
space $\mathcal {H}_{A}$ and introduce an assistant state
\begin{eqnarray}
|\Lambda\rangle&=&\alpha_{1}|1\rangle_{a} |\phi_{1}\rangle_{AB}+
\alpha_{2}|2\rangle_{a}|\phi_{2}\rangle_{AB}\nonumber\\
& &+\cdot\cdot\cdot+\alpha_{n}|n\rangle_{a}|\phi_{n}\rangle_{AB},
\label{assis-state}
\end{eqnarray}
where the subscripts $a$ denote that
$|i\rangle,i=1,2,\cdot\cdot\cdot,n$ are states in Hilbert space
$\mathcal {H}_{a}$ and the subscripts $AB$ denote that
$\phi_1,\phi_2\cdot\cdot\cdot\phi_n$ are bipartite states in Hilbert
space $\mathcal {H}_{A}\bigotimes\mathcal {H}_{B}$. In fact, we can
also think that $|\Lambda\rangle$ is a tripartite quantum state of
three systems in the Hilbert space $\mathcal
{H}_{a}\mathcal\bigotimes\mathcal {H}_{A}\bigotimes\mathcal
{H}_{B}$, the two elucidations is equivalent. In the following text
we will omit these subscripts. We request that $|i\rangle$'s
($i=1,2\cdot\cdot\cdot,n$) are mutual orthogonal and normalized, so
they become a base of $\mathcal {H}_{a}$. In addition, we require
that $\sum\limits_{i=1}^{n}{|\alpha_{i}|^2}=1$ and
 $\phi_1,\phi_2,\cdot\cdot\cdot,\phi_n$ are all normalized, so
$|\Lambda\rangle$ is normalized too. Bob's reduced state for $|\Lambda\rangle$ is
\begin{eqnarray}
\rho_{B}=\sum\limits_{i=1}^{n}{|\alpha_{i}|^2\tr_{A}(|\phi_{i}\rangle\langle\phi_{i}|)}.
\end{eqnarray}
By using inequality (\ref{assis-ineq}), we have
\begin{eqnarray}
S(\rho_{B})&\leq&\sum\limits_{i=1}^{n}{|\alpha_{i}|^2S(\tr_{A}
(|\phi_{i}\rangle\langle\phi_{i}|))}\nonumber\\
&&+(-\sum\limits_{i=1}^{n}{|\alpha_{i}|^2\log{|\alpha_{i}|^2}}).
\label{rhob1}
\end{eqnarray}
Now we introduce another normalized and orthogonal base
\{$|\xi_{i}\rangle,i=1,2,\cdot\cdot\cdot,n$\} in Hilbert space
$\mathcal {H}_{a}$, and adopt it to express base $\{|i\rangle\}$ as
\begin{eqnarray}
|1\rangle&=&{1\over N_{1}}(|\xi_{1}\rangle+|\xi_{2}\rangle)\nonumber\\
|2\rangle&=&{1\over N_{2}}(|\xi_{1}\rangle-|\xi_{2}\rangle+|\xi_{3}\rangle)\nonumber\\
|3\rangle&=&{1\over N_{3}}(|\xi_{1}\rangle-|\xi_{2}\rangle-2|\xi_{3}\rangle+|\xi_{4}\rangle)\nonumber\\
& &\cdot\nonumber\\
& &\cdot\nonumber\\
|i\rangle&=&{1\over N_{i}}[N_{i-1}|i-1\rangle-|\xi_{i}\rangle-({N_{i-1}}^2-1)|\xi_{i}\rangle\nonumber\\
&&+|\xi_{i+1}\rangle]\nonumber\\
& &\cdot\nonumber\\
& &\cdot\nonumber\\
|n-1\rangle&=&{1\over N_{n-1}}[N_{n-2}|n-2\rangle-|\xi_{n-1}\rangle\nonumber\\
&&-({N_{n-2}}^2-1)|\xi_{n-1}\rangle+|\xi_{n}\rangle]\nonumber\\
& &\nonumber\\
|n\rangle&=&{1\over N_{n}}[N_{n-1}|n-1\rangle-|\xi_{n}
\rangle-({N_{n-1}}^2-1)|\xi_{n}\rangle]. \nonumber\\
\label{base-trans}
\end{eqnarray}
These $N_{i}$'s are $|i\rangle$'s normalization coefficients. They
are all positive. It can be easily found that the orthogonality of
${|i\rangle}$'s is preserved. For a given $n$, using followed
expressions we can easily calculated all $N_{i}$'s,
\begin{eqnarray}
&&{N_{1}}^2=2\nonumber\\
&&{N_{j}}^2=\prod\limits_{i=1}^{j-1}{{N_{i}}^2}+1\ \ \ \ \   1<j<n\nonumber\\
&&{N_{n}}^2=\prod\limits_{i=1}^{n-1}{{N_{i}}^2}.
\label{N}
\end{eqnarray}

From Eqs. (\ref{assis-state}) and (\ref{base-trans}), we have
\begin{widetext}
\begin{eqnarray}
|\Lambda\rangle&=&\alpha_{1}{1\over N_{1}}(|\xi_{1}\rangle+|\xi_{2}\rangle)|\phi_{1}\rangle
+\alpha_{2}{1\over N_{2}}(|\xi_{1}\rangle-|\xi_{2}\rangle+|\xi_{3}\rangle)|\phi_{2}\rangle
+\cdot\cdot+\alpha_{i}{1\over N_{i}}[|\xi_{1}\rangle-\cdot\cdot\cdot-({N_{i-1}}^2-1)|\xi_{i}\rangle
+|\xi_{i+1}\rangle]|\phi_{i}\rangle\nonumber\\
&&+\cdot\cdot+\alpha_{n}{1\over N_{n}}[|\xi_{1}\rangle-\cdot\cdot\cdot-({N_{n-1}}^2-1)|\xi_{n}\rangle]|\phi_{n}\rangle\nonumber\\
&=&\left({\alpha_{1}\over N_{1}}|\phi_{1}\rangle+{\alpha_{2}\over
N_{2}}|\phi_{2}\rangle+\cdot\cdot\cdot+ {\alpha_{n}\over
N_{n}}|\phi_{n}\rangle\right)|\xi_{1}\rangle+|C_{2}\rangle|\xi_{2}\rangle+\cdot\cdot\cdot+|C_{n}\rangle
|\xi_{n}\rangle,
\end{eqnarray}
where $|C_{i}\rangle$'s are some superposition states of
$\phi_1,\phi_2,\cdot\cdot\cdot,\phi_n$. We do not present their
explicit expressions here because they will not be requested below.
It should be noted that $(\frac{\alpha_{1}}
{N_{1}}|\phi_{1}\rangle+\frac{\alpha_{2}}
{N_{2}}|\phi_{2}\rangle+\cdot\cdot\cdot+ \frac{\alpha_{n}}
{N_{n}}|\phi_{n}\rangle)$ and $|C_{i}\rangle$'s are all not
normalized, so we can write $|\Lambda\rangle$ as
\begin{eqnarray}
|\Lambda\rangle&=&\left\|{\alpha_{1}\over
N_{1}}|\phi_{1}\rangle+{\alpha_{2}\over
N_{2}}|\phi_{2}\rangle+\cdot\cdot\cdot+ {\alpha_{n}\over
N_{n}}|\phi_{n}\rangle\right\|\cdot\left( {({\alpha_{1}\over
N_{1}}|\phi_{1}\rangle +{\alpha_{2}\over
N_{2}}|\phi_{2}\rangle+\cdot\cdot\cdot+ {\alpha_{n}\over
N_{n}}|\phi_{n}\rangle)\over \|{\alpha_{1}\over
N_{1}}|\phi_{1}\rangle +{\alpha_{2}\over
N_{2}}|\phi_{2}\rangle+\cdot\cdot\cdot+
{\alpha_{n}\over N_{n}}|\phi_{n}\rangle\|}\right)|\xi_{1}\rangle\nonumber\\
&&+\||C_{2}\rangle\|{|C_{2}\rangle\over
\||C_{2}\rangle\|}|\xi_{2}\rangle
+\cdot\cdot\cdot+\||C_{n}\rangle\|{|C_{n}\rangle\over
\||C_{n}\rangle\|}|\xi_{n}\rangle . \label{answer-one}
\end{eqnarray}
Using the normalization of the $|\Lambda\rangle$ and the
orthogonality of the $|\xi_{n}\rangle$'s, we obtain
\begin{eqnarray}
\left\|{\alpha_{1}\over N_{1}}|\phi_{1}\rangle+{\alpha_{2}\over
N_{2}}|\phi_{2}\rangle+\cdot\cdot\cdot+ {\alpha_{n}\over
N_{n}}|\phi_{n}\rangle\right\|^2+\sum\limits_{i=2}^{n}{\||C_{i}\rangle\|^2}
=1 . \label{normalized}
\end{eqnarray}
From Eq. (\ref{answer-one}) we obtain another expression of Bob's
reduced state
\begin{eqnarray}
\rho_{B}&=&\tr_{A}(|\Lambda\rangle\langle\Lambda|)\nonumber\\
&=&\left\|{\alpha_{1}\over N_{1}}|\phi_{1}\rangle+{\alpha_{2}\over
N_{2}}|\phi_{2}\rangle+\cdot\cdot\cdot+
{\alpha_{n}\over N_{n}}|\phi_{n}\rangle\right\|^2\nonumber\\
&&\times\tr_{A}\left[{\left({\alpha_{1}\over N_{1}}|\phi_{1}\rangle
+{\alpha_{2}\over N_{2}}|\phi_{2}\rangle+\cdot\cdot\cdot+
{\alpha_{n}\over
N_{n}}|\phi_{n}\rangle\right)\left({\alpha_{1}^{\ast}\over
N_{1}}\langle\phi_{1}| +{\alpha_{2}^{\ast}\over
N_{2}}\langle\phi_{2}|+\cdot\cdot\cdot+ {\alpha_{n}^{\ast}\over
N_{n}}\langle\phi_{n}|\right)\over \left\|{\alpha_{1}\over
N_{1}}|\phi_{1}\rangle+{\alpha_{2}\over
N_{2}}|\phi_{2}\rangle+\cdot\cdot\cdot+
{\alpha_{n}\over N_{n}}|\phi_{n}\rangle\right\|^2}\right]\nonumber\\
&&+\||C_{2}\rangle\|^2\tr_{A}\left({|C_{2}\rangle\langle C_{2}|\over
\||C_{2}\rangle\|^2}\right)+\cdot\cdot\cdot+\||C_{n}\rangle
\|^2\tr_{A}\left({|C_{n}\rangle\langle C_{n}|\over \||C_{n}\rangle
\|^2}\right). \label{rhob2}
\end{eqnarray}
From Eqs. (\ref{assis-ineq}), (\ref{rhob1}), (\ref{rhob2}), and
(\ref{normalized}), we derive
\begin{eqnarray}
&&\left\|{\alpha_{1}\over N_{1}}|\phi_{1}\rangle+{\alpha_{2}\over
N_{2}}|\phi_{2}\rangle+\cdot\cdot\cdot+
{\alpha_{n}\over N_{n}}|\phi_{n}\rangle\right\|^2\nonumber\\
&&\times S\left(\tr_{A}\left[{\left({\alpha_{1}\over
N_{1}}|\phi_{1}\rangle +{\alpha_{2}\over
N_{2}}|\phi_{2}\rangle+\cdot\cdot\cdot+ {\alpha_{n}\over
N_{n}}|\phi_{n}\rangle\right)\left({\alpha_{1}^{\ast}\over
N_{1}}\langle\phi_{1}| +{\alpha_{2}^{\ast}\over
N_{2}}\langle\phi_{2}|+\cdot\cdot\cdot+ {\alpha_{n}^{\ast}\over
N_{n}}\langle\phi_{n}|\right)\over \left\|{\alpha_{1}\over
N_{1}}|\phi_{1}\rangle+{\alpha_{2}\over
N_{2}}|\phi_{2}\rangle+\cdot\cdot\cdot+
{\alpha_{n}\over N_{n}}|\phi_{n}\rangle\right\|^2}\right]\right)\nonumber\\
&&+\||C_{2}\rangle\|^2S\left(\tr_{A}\left({|C_{2}\rangle\langle
C_{2}|\over \||C_{2}\rangle\|^2}\right)\right)+\cdot\cdot\cdot
+\||C_{n}\rangle\|^2S\left(\tr_{A}\left({|C_{n}\rangle\langle
C_{n}|\over \||C_{n}\rangle
\|^2}\right)\right)\nonumber\\
&\leq&S(\rho_{B})\leq\sum\limits_{i=1}^{n}{|\alpha_{i}|^2S(\tr_{A}(|\phi_{i}\rangle\langle\phi_{i}|))}
+\left( -\sum\limits_{i=1}^{n}{|\alpha_{i}|^2\log{|\alpha_{i}|^2}}
\right). \label{answer-two}
\end{eqnarray}
Since all
$\||C_{i}\rangle\|^2S\left(\tr_{A}\left(\frac{|C_{i}\rangle\langle
C_{i}|}{ \||C_{i}\rangle\|^2}\right)\right)\geq0$, from Eq.
(\ref{answer-two}) we have
\begin{eqnarray}
&&\left\|{\alpha_{1}\over N_{1}}|\phi_{1}\rangle+{\alpha_{2}\over
N_{2}}|\phi_{2}\rangle+\cdot\cdot\cdot+
{\alpha_{n}\over N_{n}}|\phi_{n}\rangle\right\|^2\nonumber\\
&&\times S\left(\tr_{A}\left[{\left({\alpha_{1}\over
N_{1}}|\phi_{1}\rangle +{\alpha_{2}\over
N_{2}}|\phi_{2}\rangle+\cdot\cdot\cdot+ {\alpha_{n}\over
N_{n}}|\phi_{n}\rangle\right)\left({\alpha_{1}^{\ast}\over
N_{1}}\langle\phi_{1}| +{\alpha_{2}^{\ast}\over
N_{2}}\langle\phi_{2}|+\cdot\cdot\cdot+ {\alpha_{n}^{\ast}\over
N_{n}}\langle\phi_{n}|\right)\over \left\|{\alpha_{1}\over
N_{1}}|\phi_{1}\rangle+{\alpha_{2}\over
N_{2}}|\phi_{2}\rangle+\cdot\cdot\cdot+
{\alpha_{n}\over N_{n}}|\phi_{n}\rangle\right\|^2}\right]\right)\nonumber\\
&\leq&\sum\limits_{i=1}^{n}{|\alpha_{i}|^2S(\tr_{A}(|\phi_{i}\rangle\langle\phi_{i}|))}
+(-\sum\limits_{i=1}^{n}{|\alpha_{i}|^2\log{|\alpha_{i}|^2}}).
\label{answer-three}
\end{eqnarray}
\end{widetext}
Defining $\alpha_{i}^{'}=\frac{\alpha_{i}}{ N_{i}}$ and noting
$\sum\limits_{i=1}^{n}{|\alpha_{i}|^2}=1$, one has $\sum\limits_{i =
1}^n {N_i ^2 |\alpha _{i}^{'} |^2 }  = 1$. Using $\alpha_{i}^{'}$ to
express Eq. (\ref{answer-three}), we finally deduce the inequality
(\ref{xymain-ineq}). When $n=2$, from (\ref{N}) we learn that
$N_{1}^{2}=N_{2}^{2}=2$ and the inequality (\ref{xymain-ineq})
reduces to the result of \cite{linden}.

It should be noticed that
$E(\alpha_1\phi_1+\alpha_2\phi_2+\cdot\cdot\cdot+\alpha_n\phi_n)\neq
E(\alpha_{1}^{'}\phi_1+\alpha_{2}^{'}\phi_2+\cdot\cdot\cdot+\alpha_{n}^{'}\phi_n)$
except the case of $n=2$, the reason is that usually
$(|\alpha_{i}^{'}|^2/ |\alpha_{j}^{'}|^2)=(|\alpha_{i}|^2/
|\alpha_{j}|^2) \times ({N_{j}}^2/ {N_{i}}^2)\neq(|\alpha_{i}|^2/
|\alpha_{j}|^2)$.

In the case that $\phi_1,\phi_2,\cdot\cdot\cdot,\phi_n$ are
orthogonal but not biorthogonal, for deducing the inequality
(\ref{xymain-ineq}) we only need to replace
$\|\alpha_1|\phi_1\rangle+\alpha_2|\phi_2\rangle+\cdot
\cdot\cdot+\alpha_n|\phi_n\rangle\|^2$ with $
(|\alpha_{1}|^2+|\alpha_{2}|^2+\cdot\cdot\cdot+|\alpha_{n}|^2)$.

In the expression (\ref{3}) and the inequality (\ref{xymain-ineq}),
the constrained condition $\sum\limits_{i = 1}^n {N_i ^2 |\alpha _i
|^2 }  = 1$ obviously makes the function $h_n$ positive, and does
not add any restrictions on the superposition itself. For example,
for any superposition state
$|\Psi\rangle=\alpha_1|\phi_1\rangle+\alpha_2|\phi_2\rangle+
\cdot\cdot\cdot+\alpha_n|\phi_n\rangle$, multiplying it by a
constant $\frac{1}{\sqrt{\sum\limits_{i = 1}^n {N_i ^2 |\alpha _i
|^2 }}}$, we have a new state
${|\Psi\rangle}^{'}=\alpha_{1}^{'}|\phi_1\rangle+\alpha_{2}^{'}|\phi_2
\rangle+\cdot\cdot\cdot +\alpha_{n}^{'}|\phi_n\rangle$. These
$\alpha_{i}^{'}$'s satisfy $\sum\limits_{i = 1}^n {N_i ^2 |\alpha
_{i}^{'} |^2 }  = 1$. Recalling that $E(\Psi)$ in inequality
(\ref{xymain-ineq}) denotes the entanglement of the normalized
version of state $\Psi$, $E(\Psi)=E(\Psi^{'})$ and we can use
inequality (\ref{xymain-ineq}) to discuss arbitrary superposition
states. It is easy to find that for arbitrary
$|\Psi\rangle=\alpha_1|\phi_1\rangle+\alpha_2|\phi_2\rangle+
\cdot\cdot\cdot+\alpha_n|\phi_n\rangle$ without the constrained
condition $\sum\limits_{i = 1}^n {N_i ^2 |\alpha _i |^2 }=1$, we
have an inequality
\begin{widetext}
\begin{eqnarray}
\|\alpha_1|\phi_1\rangle+\alpha_2|\phi_2\rangle+\cdot\cdot
\cdot+\alpha_n|\phi_n\rangle\|^2 \cdot
E(\alpha_1\phi_1+\alpha_2\phi_2+\cdot\cdot\cdot+\alpha_n\phi_n)\leq
\sum\limits_{i = 1}^n {N_i ^2 |\alpha _i |^2 E(\phi_i)}+h_{n}^{'},
\label{ineq01}
\end{eqnarray}

where $h_{n}^{'}=-\sum\limits_{i = 1}^n {N_i ^2 |\alpha _i |^2
\log(N_i ^2 |\alpha _i |^2)}+\log(\sum\limits_{i = 1}^n {N_i ^2
|\alpha _i |^2})\cdot\sum\limits_{i = 1}^n {N_i ^2 |\alpha _i |^2}$.

In fact, if we multiply $\ket{\Psi}$ by a different constant
$\frac{1}{\sqrt{\sum\limits_{i = 1}^n {N_{i} ^{'2} |\alpha _i |^2
}}}$, where $\{N^{'}_{i}\}$ is a different order of $\{N_{i}\}$, we
have the same state $\ket{\Psi}^{'}$ and a different constrained
condition $\sum\limits_{i = 1}^n {N_i ^{'2} |\alpha _{i}^{'} |^2 }
= 1$, as a consequence we will obtain a different inequality
(\ref{ineq01}). Considering this effect, for any superposition
state, the inequality (\ref{ineq01}) should be moderated as
\begin{eqnarray}
\|\alpha_1|\phi_1\rangle+\alpha_2|\phi_2\rangle+\cdot\cdot
\cdot+\alpha_n|\phi_n\rangle\|^2 \cdot
E(\alpha_1\phi_1+\alpha_2\phi_2+\cdot\cdot\cdot+\alpha_n\phi_n)\leq
\min_{\{N_{i}\}}\left\{ \sum\limits_{i = 1}^n {N_i ^2 |\alpha _i |^2
E(\phi_i)}+h_{n}^{'} \right\}, \label{ineq02}
\end{eqnarray}
\end{widetext}
where $\min_{\{N_{i}\}}$ means that the lowest bound is taken over
all possible sets of $\{ N_i\}$, associated with different values of
$\alpha_i$ and $\phi_i$ in the sum and $h_n^{'}$ on the right-hand
side of Eq. (\ref{ineq02}).

In summary, we present an upper bound on the entanglement of
superposition states with more than two components. The bound
expressed in inequality (\ref{xymain-ineq}) with arbitrary $\{
N_i\}$ most likely not be the best one. For many cases, we find that
the bound is loose. We suppose that there may be two reasons for the
looseness of the bound: (i) In the derivation of inequality
(\ref{xymain-ineq}) we have used the relation $\sum\limits_{i=1}^n
{p_i S(\rho_i)}\leq\sum\limits_{i=1}^n {p_i S(\rho_i)}+H$ and so the
difference between conditions for two equalities Eq.
(\ref{assis-ineq}) should make the bound hard to be achieved. (ii)
In the derivation of Eq. (\ref{answer-three}), we have dropped the
terms such as
$\||C_{i}\rangle\|^2S\left(\tr_{A}\left(\frac{|C_{i}\rangle\langle
C_{i}|}{ \||C_{i}\rangle\|^2}\right)\right)$, and this may reduce
the value of the left-hand side in inequality (\ref{xymain-ineq}).
In \cite{gour}, the author has shown that the bound in \cite{linden}
is not tight and given a tighter upper bound for the case of
superposition with two components. In the case of $n=2$, our
inequality (\ref{xymain-ineq}) reduces to the result of
\cite{linden}, so the result of Gour \cite{gour} implies that there
may be a better inequality for the entanglement of superposition
with more than two components. This may be a task for future work.

{\it Acknowledgments} We would like to thank the referee who
reminded us to note the effect which result in Eq. (\ref{ineq02}).
This work was supported by National Foundation of Natural Science in
China Grant Nos. 60676056 and 10474033, and by the China State Key
Projects of Basic Research (2005CB623605 and 2006CB0L1000).


%
%
%
%
%
%
%

\end{document}